# Passively Q-switched Erbium-doped fiber laser using $Fe_3O_4$-nanoparticle saturable absorber

Xuekun Bai,[1,a] Chengbo Mou,[2,a] Luxi Xu,[1] Sujuan Huang,[1] Tingyun Wang,[1] Shengli Pu,[3]

Xianglong Zeng[1,*]

[1]*The Key Lab of Specialty Fiber Optics and Optical Access Network, Shanghai University, 200072 Shanghai, China*
[2] *Aston Institute of Photonic Technologies (AIPT), Aston University, Aston Triangle, Birmingham B4 7ET, United Kingdom*
[3] *College of science, University of Shanghai for Science and Technology, 200093 Shanghai, China*
[a]*These authors contribute equally to the work*
*\*Corresponding author: zenglong@shu.edu.cn*

We experimentally demonstrate a passively Q-switched erbium-doped fiber laser (EDFL) operation by using a saturable absorber based on $Fe_3O_4$ nanoparticles (FONP) in magnetic fluid (MF). As a kind of transition metal oxide, the FONP has a large nonlinear optical response with a fast response time for saturable absorber. By depositing MF at the end of optical fiber ferrule, we fabricated a FONP-based saturable absorber, which enables a strong light-matter interaction owing to the confined transmitted optical field within the single mode fiber. Because of large third-order optical nonlinearities of FONP-based saturable absorber, large modulation depth of 8.2% and non saturable absorption of 56.6% are demonstrated. As a result, stable passively Q-switched EDFL pulses with maximum output pulse energy of 23.76 nJ, repetition rate of 33.3 kHz, and pulse width of 3.2 μs are achieved when the input pump power is 110 mW at the wavelength of 980 nm. The laser features a low threshold pump power of ~15 mW.

*OCIS codes:* (140.3540) Lasers, Q-switched; (160.2260) Ferroelectrics; (160.4236) Nanomaterials; (160.6000) Semiconductor materials.

## 1. INTRODUCTION

Q-switched erbium-doped fiber lasers (EDFLs) are useful light sources due to their potential applications in remote sensing, range finding, laser processing, optical communications, military and etc. [1–2]. Compared with the active Q-switching in the laser cavities, passive Q-switching based on intensity saturable absorbers (SAs) possesses the advantages of compactness, low-cost, and simple cavity configuration. It has been reported that various kinds of functional materials are used as SAs to achieve passive Q-switching, such as graphene [3], gold nanorods [4–5], carbon nanotubes [2, 6], $Bi_2Se_3$ [7] , $MoS_2$ [8–10], $WS_2$ [11–12], and so on. Besides, exploring other new SA materials and designing new schemes of passive Q-switching EDFLs is still attractive.

$Fe_3O_4$ nanoparticles (FONP) show remarkable interesting phenomena including super-paramagnetism, high field irreversibility, and extra anisotropy contributions because of their finite size and surface effects [13]. FONP illustrates great potential in a variety of applications including medical applications [14], microwave devices [15], magneto-optics devices [16–19], and magnetic field sensors [20–21]. Furthermore, FONP exhibits nonlinear photonic properties, such as two-photon absorption [22], optical limiting [23], and nonlinear scattering [24]. As a kind of transition metal oxide, FONP shows large third-order optical nonlinearities susceptibility $\chi^{(3)}$ indicating a large nonlinear optical response for many applications [25–27]. And the recovery time of FONP was evaluated as 1-30 ps [28]. Recently, the energy band gap of FONP was observed so that such materials can be classified as a type of semiconductor [29–30]. The band gap energy of FONP can be tuned by changing the diameter of the nanoparticles [28], which is finely controlled by adjusting the pH and ionic strength of the precipitation medium in the process of preparation [13, 30]. Therefore, FONP is claimed to be a novel type of promising nonlinear optical material, which may find wide potential applications in photonics.

In this work, we propose and demonstrate an all-fiber passively Q-switched EDFL by using FONP-based nonlinear optical SA. FONP is deposited on the optical fiber end from the commercial magnetic fluid (MF), which does not need any additional pre-preparation. The fabrication of FNOP-based SA device is simple and straightforward. Compared to other types of SA, the feature of controllable band energy is highlighted in the FONP-based device that plays key role in implementing SA with predefined and optimized saturation intensity, operation bandwidth and non-bleachable loss.

## 2. EXPERIMENTS AND RESULTS

FONP used in our work is derived from the commercial water-based MF (Ferrotec, EMG 708), which is usually synthesized by chemical coprecipitation method [31]. The surfactant used in this MF is anionic, which prevents the FONP from agglomerating due to van der Waals attraction. The average diameter of original FONP in MF is about 10 nm. The sizes of these FONP are indeed very small so that thermal energies are comparable to their gravitational forces. Thus the sedimentations can be avoided and the water based MF is a stable colloidal system. When the MF is exposed in air at room temperature, the stable colloidal system is transformed to a solid-state thin film in about 10 minutes. This is attributed to the unsaturated fatty acid based anionic surfactant and the evaporation of water carrier liquid. These facilitate the fabrication of FONP-based SA with MF.

To prepare the FONP-based SA, firstly the optical fiber ferrule end is carefully immersed into a single drop of MF, and then the MF is transferred to the fiber ferrule end due to the surface tension of aqueous liquid. Next, the ferrule with MF is exposed to air at room temperature for 30 minutes to allow the maximum evaporation of the water carrier in MF. Then the fabrication of FONP-based SA is finished. Fig. 1 (a) shows the SEM image of the cross section of optical fiber connector with dried MF film. The thickness of the film deposited on the top of optical fiber connector is ~20 μm. The SEM image of the FNOP in dried MF film is shown in Fig. 1 (b). The diameter of FNOP is non-uniform and the measured average diameter is about 50 nm. This is due to the random agglomeration of FONP during the evaporation process of water when MF is exposed in air at room temperature.

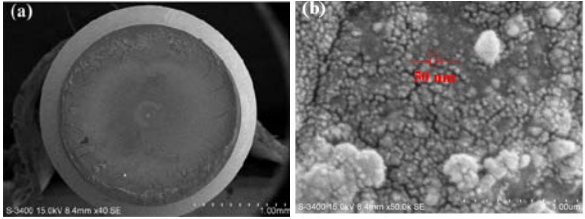

Fig. 1. SEM image of (a) FONP-deposited optical fiber ferrule end and (b) FONP in the film.

To investigate the FONP-based SA, the as-fabricated SA is incorporated into the laser cavity using the popular sandwich structure via another standard fiber connector. The measured optical absorption of the FONP-based SA is illustrated in Fig. 2 (a). We can see that the FONP-based SA has a broad absorption band across the infrared band from 1438 nm to 1628 nm. The linear absorption of the FONP-based SA is ~63% at the wavelength of 1560 nm. The nonsaturable loss can be reduced effectively by optimizing the thickness and uniformity of FONP film, or using tapered or polished fiber covered FONP. Because the finite thickness of the SA film between the two optical fiber ferrules, interference pattern is seen clearly from the measured absorption spectrum. The visibility of SA-induced interference is relatively small (less than 5%), which can be neglected in the fiber laser cavity. The measured insertion loss of SA is 4.1 dB.

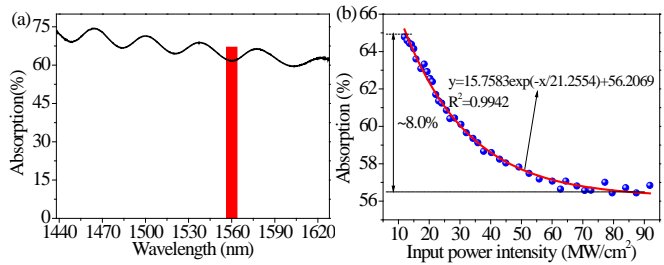

Fig. 2. (a) Measured linear absorption and (b) saturable absorption of the FONP-based SA device.

The balanced twin-detector measurement is performed to investigate the absorption property of the FONP-based SA [10]. The measurement system consists of a home-made femtosecond fiber laser as light source (center wavelength: 1560 nm, repetition rate: 40 KHz, 10 ps pulse duration, 10 mW output), a variable optical attenuator, a 3-dB optical coupler and a two-channel power meter. The saturable absorption (modulation depth) and the non-saturable absorption of the FONP-based SA is measured to be about 8.2% and 56.6%, respectively, as shown in Fig. 2 (b). And the order of this saturable absorption of -0.0781 is obtained from the saturable absorption fitting with a power law function $x^b$. The saturable absorption of FONP could be contributed to the combined effect of third-order nonlinear absorption and nonlinear scattering of FONP [24, 26, 27].

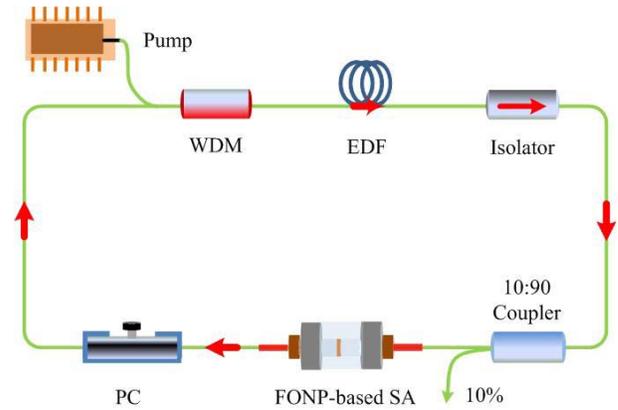

Fig. 3. Schematic of the passively Q-switched EDFL with the FONP-based SA. The inset is the fabricated FONP-based SA.

Figure 3 shows the schematic configuration of the passively Q-switched EDFL incorporating the FONP-based SA. The red arrow indicates the light propagation direction in the laser cavity. A length of 7.5 m EDF (NUFERN, EDFC-980-HP) is pumped by a 980 nm laser diode through a 980/1550 nm wavelength division multiplexing (WDM) coupler. The EDF used in this experiment has a nominal absorption of 6.0 ±1.0 dB/m at 1530 nm with group velocity dispersion (GVD) of +15.5 $ps^2$/km at 1550 nm. The pigtail fiber (Corning, HICER98) of the WDM couplers is ~1.86 m long with a dispersion of -0.25 $ps^2$/km at 1550 nm. An in-fiber polarization-independent isolator is utilized to force the unidirectional operation of the laser cavity. At the laser output, 10% of the output power is extracted from the laser cavity by a 90/10 coupler. The rest of the fibers in the laser ring cavity are standard single-mode fibers (SMFs) with a typical GVD of -23 $ps^2$/km at 1550 nm. The total length of the cavity is about 19.4 m, and the overall dispersion of the cavity is about -0.1 $ps^2$ at 1550 nm. A polarization controller (PC) is used in the cavity to optimize the cavity birefringence. The temporal and spectral characteristics of the Q-switched fiber laser output are recorded by a 10 GHz photo detector (CONQUER, KG-PD-10G-FP) followed by a 300 MHz oscilloscope (Tektronix, DPO3032) and an optical spectrum analyzer (YOKOGAWA, AQ6370C).

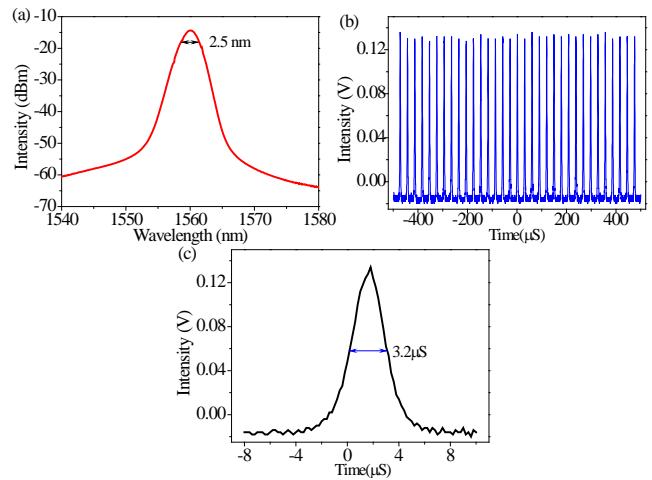

Fig. 4. (a) The spectrum; (b) the pulse train; and (c) the profile of the Q-switched pulses.

Passively Q-switching of the constructed fiber laser self-starts at pump power of 15 mW. Compared with the reported works [5,

12], the threshold power of this laser is relatively low, which is attributed to the larger third-order optical nonlinearities induced by the FONP. The 3 dB bandwidth of the optical spectrum is measured to be 2.5 nm at the center wavelength of 1560 nm when pumping at 110 mW as shown in Fig. 4 (a). It is relative broad compared with the reported results [3, 7, 10]. This is duo to the $Fe_3O_4$ nanoparticles induced scattering effects [24]. The measured average output power is ~0.8 mW using a commercial power meter (Yiai, AV6334). The corresponding output pulse train with a repetition rate of 33.3 kHz and the measured pulse duration of 3.2 μs are shown in Fig. 4 (b) and (c), respectively.

## 3. DISCUSSIONS

Figure 5 (a) shows spectra of Q-switched output pulse with different pump powers. By increasing the pump power from 15 mW to 110 mW, the 3 dB bandwidth of the spectrum increases from 0.38 nm to 2.5 nm as shown in Fig. 5 (b). And the relationship between 3dB bandwidth and pump power can be fitted with y=0.1235+0.02285x, and the corresponding fitting coefficient $R^2$ is 0.9327. In the experiment, when more gain is provided to saturate the SA, typical Q-switching behavior has been observed. Pulse energy is found to change with the increment of the pump power and the output power monotonically. The corresponding output power increased from 94 μW to 794 μW and the single pulse energy varied from 12.13 nJ to 23.76 nJ, as shown in Fig.5 (c). The laser efficiency is about 0.7%. The output pulse with energy of 23.76 nJ was achieved at the maximum pump power of 110 mW.

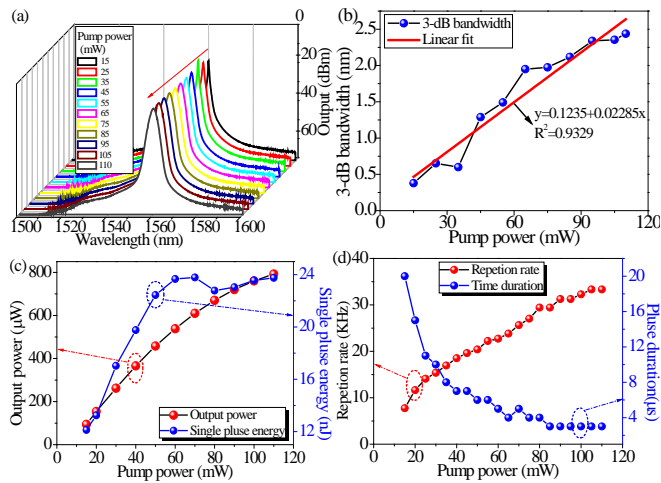

Fig. 5. (a) The output spectrum; (b) the corresponding 3 dB bandwidth; (c) the average output power and the single pulse energy; (d) the repetition rate and pulse width variation with the increasing pump power.

During the process of increasing pump powers, the pulse duration of output stable pulses reduces from 20.0 μs to 3.2 μs while the repetition rate monotonically increases from 7.8 kHz to 33.3 kHz, as shown in Fig. 5 (d). It can be found that the pulse duration becomes smaller and the repetition rate becomes larger with the increasing pump power. This also presents a typical feature of passively Q-switched lasers. The pump rate for the upper laser level increases with increasing the pump power and causes the reduction of the pulse width and the increase of the repetition rate [2, 4]. The demonstrated laser shows no obvious degradation at the laboratory condition for 2 hours. In our experiment, no stable pulse trains are observed when the pump power goes beyond 110 mW. The laser becomes CW operation when the pump power increases from 110 mW to 600 mW. If the pump power is reduced to 110 mW again, the Q-switching reappeared. This is because FONP exhibits a saturable absorption at moderate laser intensities, while optical limiting induced by excited states absorption would occur at higher intensities [23, 28, 32].

To verify the effects of the FONP-based SA on Q-switching, two optical fiber connectors connected by a flange are inserted in the EDFL cavity. But there is no FONP deposited on the end of the optical fiber connector. And the corresponding SMFs' length is not changed. Fig. 6 (a) and (b) present the measured emission spectrum of the EDFL and the laser intensity as a function of time for a pump power of 40 mW, respectively. As we can see from Fig. 6, only continuous wave laser operation is obtained and no pulsed operation was observed when the pump power is increased from 0 mW to 680 mW, which confirms that the above Q-switched EDFL has been induced by the FONP-based SA.

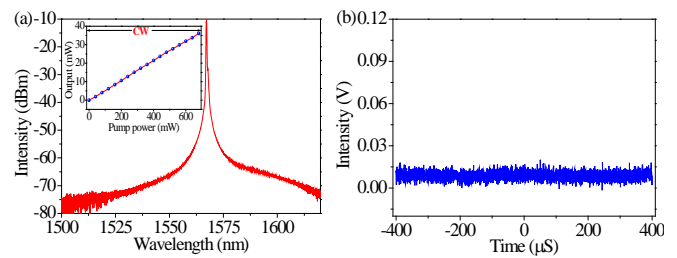

Fig. 6. (a) Emission spectrum of the EDFL without FONP-based SA and (b) laser intensity as a function of time for a pump power of 40 mW. The insert in (a) is the continuous wave laser output with the increasing pump power.

For a Q-switched fiber laser with higher pulse energy output, lower repetition rate and pump power is preferred because the intra-cavity power can be more effectively coupled to each pulse at this situation. Although such pulse energy still cannot outperform the laser using conventional semiconductor-based SAs, further improvement for scaling up pulse energy is foreseeable with such FONP-based SA by optimizing the fabrication procedure, especially the mean diameter of FONP, the thickness and the uniformity of the MF deposited on the cross section of fiber connector. Using higher gain active fiber and high output coupling ratio in the cavity may also boost the output power. The length of EDF and the cavity need to be more reasonable for improving output laser characteristics. Duo to the tunable magneto-optical properties of FONP, the Q-factor tuneability of Q-switching based on all-fiber-FONP-based SA could be controlled by an external magnetic field. The thermal effects of FONP should be avoided, which decrease the non-linear absorption of FONP-based SA. So, the amount of FONP used in experiments should be small enough and the experimental cooling should be considered carefully. The potential photonics application of FONP-based SA in the far infrared range (2 or 3 um) and mode locking by using the enhanced FONP-based SA and optimized EDFL cavity will be investigated in the future work. And in order to explore more ultrafast photonics applications, the relaxation dynamics of FONP will be studied experimentally.

## 4. CONCLUSIONS

In conclusion, we have investigated the nonlinear optical absorption of FNOP derived from MF. With the FONP-based SA, an all-fiber passively Q-switched EDFL has successfully been demonstrated. The maximum output pulse energy of the proposed fiber laser is 23.8 nJ and minimum pulse width of 3.2 μs are obtained at the repetition rate of 33.3 kHz from the laser cavity when the input pump power is 110 mW. The demonstrated laser features a low threshold due to large third-order optical response with a fast response time for the FONP-based SA. The easy fabrication, good stability and robust structure of FONP-based SA will facilitate many more potential nonlinear photonic applications, which are expected to work towards ultrafast photonics and play key role in other nonlinear photonics applications.

## ACKNOWLEDGEMENTS

This work was supported by National Natural Science Foundation of China (Grant No. 11274224). And X. Zeng acknowledges the support of the Program for Professor of Special Appointment (Eastern Scholar) at Shanghai Institutions of Higher Learning.